\documentclass[11pt]{article}
\usepackage{fullpage}

\usepackage{url}
\usepackage{color}
\usepackage{graphicx}
\DeclareGraphicsExtensions{.pdf}
\usepackage{algorithmic}
\usepackage{algorithm}
\usepackage[T1]{fontenc}
\usepackage{longtable}

\begin{document}

\title{Transparent Checkpoint-Restart for Hardware-Accelerated 3D~Graphics}

\author{
Samaneh Kazemi Nafchi\thanks{This work was partially supported
          by the National Science Foundation under Grant
	  OCI-0960978, and by a grant from Intel Corporation.}
\and
Rohan Garg\footnotemark[1]
\and
Gene Cooperman\footnotemark[1]
\and
	College of Computer and Information Science \\
	Northeastern University \\
	Boston, MA 02115 / USA \\
	samaneh@ccs.neu.edu
}

\date{}

\maketitle

\begin{abstract}
Providing fault-tolerance for long-running GPU-intensive jobs requires
application-specific solutions, and often involves saving the state of 
complex data structures spread among many graphics libraries. This work
describes a mechanism for transparent GPU-independent checkpoint-restart
of 3D~graphics. The approach is based on a record-prune-replay paradigm:
all OpenGL calls relevant to the graphics driver state are recorded;
calls not relevant to the internal driver state as of the last graphics
frame prior to checkpoint are discarded; and the remaining calls are 
replayed on restart. A previous approach for OpenGL~1.5, based on a 
shadow device driver, required more than 78,000 lines of OpenGL-specific
code. In contrast, the new approach, based on record-prune-replay, is
used to implement the same case in just 4,500 lines of code. The speed of
this approach varies between 80 per cent and nearly 100~per cent of the
speed of the native hardware acceleration for OpenGL~1.5, as measured when
running the ioquake3 game under Linux. This approach has also been extended
to demonstrate checkpointing of OpenGL~3.0 for the first time,
with a demonstration for PyMol, for molecular visualization.
\end{abstract}

\section{Introduction}
\label{sec:Introduction}

The domains of scientific visualization, medical visualization, CAD, and
virtual reality often require large amounts of time for GPU rendering.
At the high end, these computations require many computers in a cluster
for highly GPU-intensive graphics rendering~\cite{CarrollEtAl12}.
At the medium- and low-end this rendering may still require large
amounts of GPU-intensive graphics rendering, but on a single computer.
For such long-running jobs, checkpointing the state of graphics is an
important consideration.  Transparent checkpointing is strongly preferred,
since writing application-specific code to save graphics state is highly
labor-intensive and error-prone.

This work presents an approach to transparent checkpoint-restart for
GPU-accelerated graphics based on record-prune-replay.  The GPU support
is vendor-independent.  It is presented in the context of OpenGL for
Linux, but the same approach would work for other operating systems and
other graphics APIs, such as Direct3D.

Previously, Lagar-Cavilla \hbox{et~al.} presented VMGL~\cite{VMGL07}
for vendor-independent checkpoint-restart.  That pioneering work
employs the approach of a shadow device driver for OpenGL, which
shadows most OpenGL calls in order to model the OpenGL state,
and restores it when restarting from a checkpoint.
While that tour de force demonstrated feasibility for most of OpenGL~1.5,
the VMGL code to maintain the OpenGL state grew to 78,000 lines of code.

This document proposes a simpler, more maintainable alternative
approach whose implementation consists of only 4,500 lines of code,
to support OpenGL~1.5.
The approach is based on replaying a log of library calls to OpenGL and
their parameters.  The OpenGL-specific code is written as a plugin on
top of a standard checkpoint-restart package, DMTCP.  The role of DMTCP
in this work corresponds to the role of virtual machine snapshots in VMGL.

The key observation behind the {\em record-prune-replay} approach
is that in order to restore the graphics state after restart,
it suffices to replay the original OpenGL calls.  More precisely,
it suffices to replay just those OpenGL calls that are relevant to
restoring the pre-checkpoint graphics state.

For example, OpenGL provides two functions, glEnableClientState and
glDisableClientState.  Both functions take a ``capability'' parameter.
Each capability is initially disabled, and can be enabled or disabled.
At the end of a sequence of OpenGL calls, a capability is enabled
or disabled based only on which of the two functions was called last
for that capability.  So, all previous calls for the same capability
parameter can be pruned.  {\em Pruning} consists of purging all OpenGL
calls not needed to reproduce the state of OpenGL that existed
at the time of the checkpoint.

The current work is implemented as a plugin on top of
DMTCP~\cite{AnselEtAl09}.  Checkpoint-restart is the process of saving
to traditional disk or SSD the state of the processes
in a running computation, and later re-starting from stable storage.
{\em Checkpoint} refers to saving the state, {\em resume} refers to
the original process resuming computation, and {\em restart} refers
to launching a new process that will restart from stable storage.
The checkpoint-restart is {\em transparent} if no modification of the
application is required.  This is sometimes called {\em system-initiated}
checkpointing.

Finally, this work is demonstrated both for OpenGL~1.5 and for OpenGL~3.0.
This is in contrast to VMGL, which was demonstrated only for OpenGL~1.5.
The complexity of supporting OpenGL~3.0 is considerably larger than
that for OpenGL~1.5.  This is because OpenGL~3.0 introduces programmable
pipelines, and shaders.  Shaders are programs compiled from a GLSL C-like
programming language.  Shaders are then passed to the server side of
OpenGL (to the GPU).  When including the OpenGL~3.0 functions,
the size of our implementation grows from 4,500 lines of code
to 6,500 lines of code (in C and Python).

\paragraph{Organization of Paper.}
In the rest of this work, Section~\ref{sec:background} provides
background on the software libraries being used.
Section~\ref{sec:design} describes the software design,
and Section~\ref{sec:algorithm} describes the algorithm
for pruning the log.
Section~\ref{sec:comparison} compares the two approaches
of shadow drivers and record-prune-replay.
Section~\ref{sec:limitations} describes the limitations of this approach,
while Section~\ref{sec:relatedWork} presents related work.
Section~\ref{sec:experiments} describes an experimental evaluation.
Finally, Section~\ref{sec:futureWork} presents future work,
and Section~\ref{sec:conclusion} presents the conclusion.

\section{Background: the Software Stack Being Used}
\label{sec:background}

\begin{figure}
\begin{center}
\includegraphics[width=0.4\columnwidth]{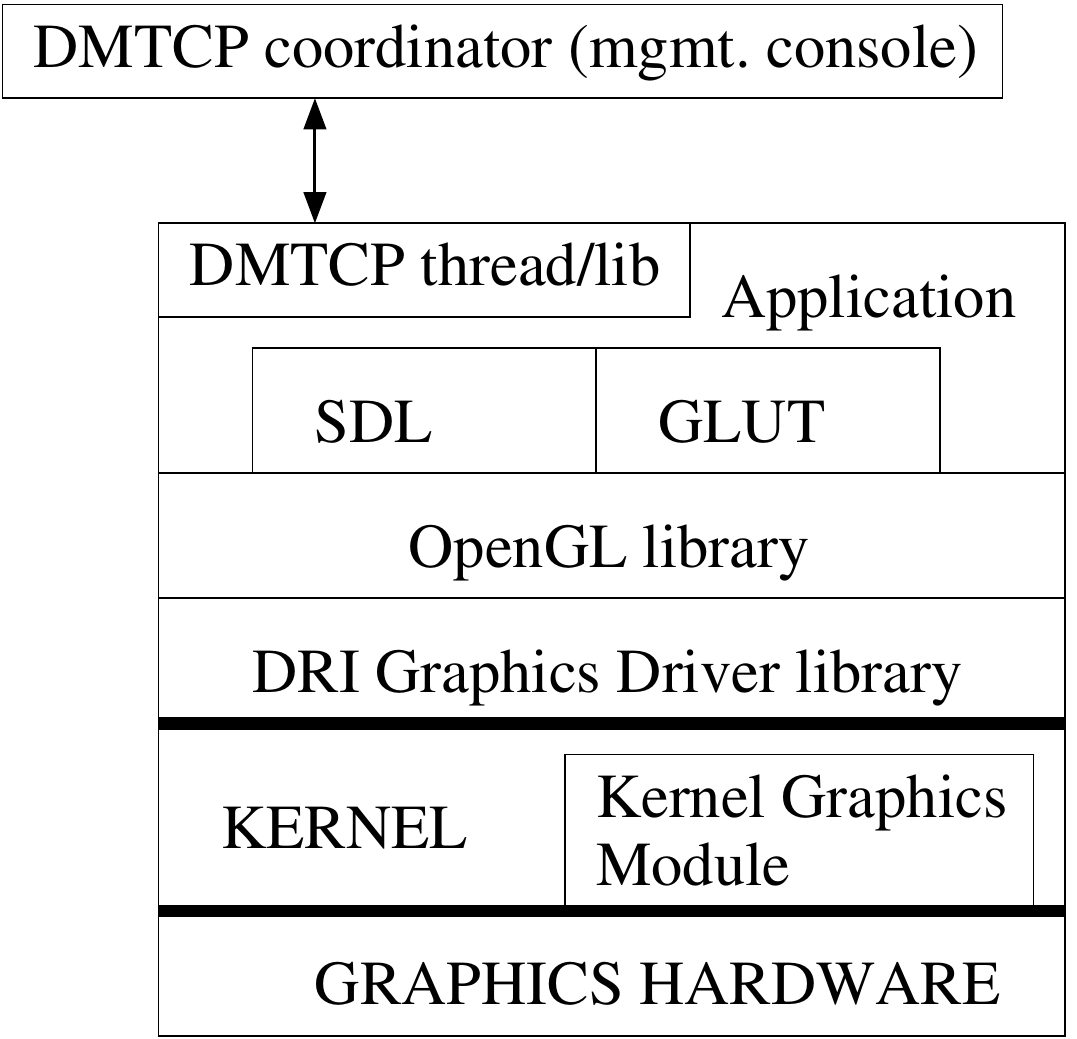}
\end{center}
\caption{\label{fig:software} Software stack of a graphics application
 to be checkpointed.}
\end{figure}

The software stack being checkpointed is described in
Figure~\ref{fig:software}.

\paragraph{Windowing Toolkits.}

OpenGL applications typically use a windowing toolkit in order to handle
external issues such as providing a graphics context based on a window
from X-Windows, and handling of input (mouse, keyboard, joystick), and
handling of sound.  Two of the more popular windowing toolkits are GLUT
and SDL.  GLUT is noted for its simplicity, while SDL is noted for its
completeness and is often used with games.  In particular, SDL is used
by ioquake3~\cite{ioquake3}, and GLUT is used by PyMol~\cite{PyMol}.
Both are built on top of lower-level GLX functions.
This work supports calls both to SDL and to GLUT, as well as to GLX.

\paragraph{DMTCP.}
\label{sec:dmtcp}
DMTCP~\cite{AnselEtAl09} (Distributed MultiThreaded CheckPointing)
provides transparent user-space checkpointing of processes.
The implementation was built using DMTCP plugins~\cite{plugin}.
In particular, two features of plugins were used:
\begin{enumerate}
\item wrapper functions around calls to the OpenGL library; and
\item event notifications for checkpoint, restart, and thread resume
during restart (PRE\_RESUME).
\end{enumerate}

Wrapper functions around the OpenGL calls were used to maintain
a log of the calls to the OpenGL library, along with the parameters
passed to that library.  At the time of restart, the log was
pruned and replayed, before returning control back to the graphics
application.

\section{Design of the DMTCP Plugin}
\label{sec:design}

As described in the introduction, the fundamental technique to be used
is record-prune-replay.  The key to pruning is to identify the dependencies
among OpenGL calls in the log.

Note that it is acceptable to over-approximate the dependencies (to
include additional dependencies that are not required), as
long as the size of the pruned log continues to remain approximately
constant over time.  This is sometimes used to produce
a simpler model of the dependencies among OpenGL calls.

In maintaining dependencies, the key observation is that if the
function~$X$ can only set the single parameter~$P$ on which $Y$
depends, then only the last call to~$X$ prior to calling~$Y$ needs
to be preserved.

However, the situation can become more complicated when a single
function~$X$ is capable of setting multiple parameters.  For example,
imagine a function: \\
\centerline{$setProperty(property\_name, property\_parameter)$.}
Imagine further that $enableProperty(property\_name)$ and
$disableProperty(property\_name)$ exist.  Then it is important the
last call be retained for each category $(fnc, arg1)$, where $fnc$
is one of the three functions $setProperty$, $enableProperty$,
and  $disableProperty$, and $arg1$ is a value of the parameter
$property\_name$.

\subsection{Design Issues}

Several additional considerations are important for a correct implementation.

\paragraph{Reinitializing the graphics library during restart.}

SDL provides a function {\tt SDL\_quit()}, that resets all state
to its initial state.  That function is required for the current
work, since at the time of restart, the pruned log of calls to
OpenGL is replayed.

GLUT does not appear to provide such a re-initialization function.
Hence, for GLUT, we make a copy of the GLUT library in memory (both text
and data segments) prior to initialization, and at the time of restart,
we overwrite this memory with a copy of the GLUT library memory as it
existed prior to the first call to that GLUT library.

Potentially, one must also worry about re-initializing a low-level
graphics library such as DRI.  We posit that this is usually not necessary,
because a well-structured graphics library like OpenGL or a generic
DRI library will determine which vendor's GPU is present, and then
directly load and initialize that vendor-specific library as part of its own
initialization.

Note that a subtle issue arises in overwriting a graphics library
during restart.  This overwriting occurs only after the
checkpoint-restart system has restarted all pre-checkpoint threads,
open files, etc.  Yet, some graphics libraries directly create
a new thread.

This issue arose, for example, in the case of the
DRI library for the ATI R300.  When it initializes itself,
it creates an auxiliary thread to maintain its internal data structures.
The auxiliary thread still exists at the time of restart, even though the
re-initialization creates an additional thread.  Hence, one must
kill the older thread on restart.  This is accomplished through
the PRE\_RESUME event notification from the DMTCP checkpoint-restart
subsystem.  Otherwise, the older thread would write to its data
structures at the same time as the newer thread, causing memory to
be corrupted.

\paragraph{SDL, GLUT and GLX.}

In this work, we intercept and log calls to SDL and GLUT as a proof
of principle.
These windowing toolkits provide high-level abstractions for calls
to GLX.
In the future, we intend to directly intercept calls to GLX in
the OpenGL library.  This will provide support for all windowing
toolkits.

\paragraph{Virtualization of Graphics IDs.}

Most of the wrappers around OpenGL functions are used only
to maintain the log.  In a few cases, we also required virtualization
of certain ids for graphics objects.  This is because the graphics id
could change between checkpoint and a restart in a new session.  Hence,
a translation table between virtual ids and real ids was maintained
for these cases.  These cases included texture ids, SDL surface ids,
and joystick ids.

\paragraph{Saving Pointers to Memory for Later Replay.}
\label{sec:pointers}

In most OpenGL functions, the entire argument is passed to the OpenGL library.
But in a few cases, a pointer to application memory is passed to the
OpenGL library.  In these cases, the data in memory at the time of
the call may be altered at the time of checkpointing.
The implementation must then copy the data into separately allocated
memory.  Examples occur for calls to {\tt glTexImage}, {\tt glTexCoordPointer},
{\tt glColorPointer}, {\tt glVertexPointer}, {\tt glBufferData},
{\tt glTexParameterfv}, and others.

\subsection{Virtualization of OpenGL ids}
\label{sec:virtualization}

Virtualization of OpenGL~ids follows a standard paradigm for DMTCP
plugins.  For each OpenGL function that refers to the type of id of
interest, the DMTCP plugin defines a wrapper function with the
same type signature as the original OpenGL function.  When the graphics
application is run under DMTCP, the wrapper function is interposed
between any calls from the graphics application to the OpenGL library.
Thus, each wrapper function is called by the application, possibly
modifies parameters of the id type of interest, and then calls
the real OpenGL function in the OpenGL library.

If the OpenGL function normally returns an id (possibly as an ``out''
parameter), then the wrapper function creates a new virtual id, enters
the (virtual,~real) pair into a translation table, and then returns
the virtual id to the application.

If the OpenGL function has an argument corresponding to that id (an
``in'' parameter), then the wrapper function assumes that the
application has passed in a virtual id.  The application uses the translation
table to replace it by the corresponding real id, before calling the real
OpenGL function in the OpenGL library.

Finally, at the time of restart, the DMTCP plugin must recreate
all graphics objects by replaying the log.  At that time, the OpenGL
library may pass back a new real id.  Since the log retained the original
virtual id, the plugin then updates the translation table by setting
the translation of the virtual id to be the new real id generated
by OpenGL.

\subsection{Representation of the OpenGL Log}
\label{sec:log}

The current implementation of the log of OpenGL calls maintains that
log as human-readable strings.  Further, the code for pruning the
log is currently a Python program.  Even worse, the log of calls is
maintained solely as a file on disk, even though this is much slower
than maintaining the log in RAM.  All of this is done for simplicity
of implementation.

While pruning the log in this way undoubtedly adds at least one order of
magnitude to the time for pruning, the effect on performance is minor.
This is because the plugin periodically creates a child process,
which prunes the log in parallel, and then notifies the parent of
success.  The child process uses an otherwise idle core of the CPU.
As Figure~\ref{fig:cdf-plot} shows, the overhead of doing this is too
small to be measured, when an idle core is available.

Nevertheless, while a human-readable log on disk is advantageous for
software development and debugging, a future version of this software
will employ C~code for pruning.  Furthermore, the current implementation
represents each OpenGL function and parameter as an ASCII string (and even
integers are written out as decimal numbers in ASCII).  The future
implementation will use binary integers to represent OpenGL function names
and most parameters.

\subsection{Support for OpenGL~3.0}
OpenGL 3.0 simplified future revisions of the API by deprecating some
older features, including glPush, glPop and display lists. Programmable
vertex and fragment processing, which was introduced in OpenGL~2.0, became
a core feature of OpenGL~3.0, and all fixed-function vertex and fragment
processing were deprecated.  The programmable pipelines have become the
standard for modern OpenGL programming.

Shaders and Vertex Buffer Objects (VBOs) are two important concepts
in modern OpenGL. These two types of objects are both stored in the
graphics memory (on the server-side). Vertex buffer objects were
introduced in OpenGL~1.5, but they had not been widely used until
OpenGL~3.0. Display lists were an earlier attempt 
to store program objects on the server-side, but using a
fixed program model in OpenGL.
These were deprecated in modern OpenGL.

Shaders are a large part of the motivation for OpenGL~3.0. A shader is a
program written in the GLSL language, which
runs on the GPU instead of the CPU. A shader program in the GLSL language
is separately compiled into binary before being used by an OpenGL program.

The plugin described here places a
wrapper function around all OpenGL functions related to shaders.
Example functions are
glCreateShader, glShaderSource, glCompileShader, glLinkProgram, and so on.
In particular, we virtualize the shader id parameter for
these functions.

Since the user application runs on the
CPU and shaders run on the GPU, a method of copying data from the CPU
to the GPU is needed. Vertex buffer objects (originally introduced
in OpenGL~1.5) fulfill this role.  This allows various types of
data (including shader programs) to
be cached in the graphics memory of the GPU.

The approach described here preserves all memory buffers associated
with a vertex buffer object by copying
them into a temporary buffer in memory.  That memory will be restored
on restart, thus allowing the plugin to copy the same buffer from
memory into graphics memory.

\subsection{Software Implementation}

We have found the log-based approach to be especially helpful from
the point of view of software engineering.  In the first stage of
implementation, we wrote wrapper functions to virtualize each of
the OpenGL ids, so as to guarantee that record-replay will work.
(In certain cases, such as testing on ioquake3, even this stage could
largely be skipped, since the application chooses its own texture and
certain other ids.)

Since the wrapper functions often follow a common pattern
(e.g., translate the id parameter from virtual to real, and then
call the real OpenGL function in the OpenGL library), we also wrote
a preprocessing language.  This language concisely takes the
declaration of an OpenGL function as input, along with a declaration
of which parameter is to be virtualized, and then generates the
desired wrapper function in~C.  Thus, full C~code is available for
compiling and debugging at all times.

However, ensuring that all pointers to memory are saved for later replay
(see Section~\ref{sec:pointers}) continued to consume significant amounts
of time for debugging.

Finally, the code for pruning the log can be developed iteratively.
Code is written for pruning a set of related OpenGL functions, and the
record-prune-replay cycle is tested for correctness by running it
inside a sophisticated OpenGL application.  Then another set of related
OpenGL functions is targeted, and similarly tested.  This breaks down the
code-writing into small, independent modules.

\section{Algorithm: Pruning the Log}
\label{sec:algorithm}

The strategy for pruning the log of OpenGL calls is based on dependencies
among OpenGL functions.  OpenGL functions for drawing a graphics frame
are identified and form the root of a dependency tree for all OpenGL
functions (along with any GLX extensions).  A node is a child of a
parent node if executing an OpenGL call of the type given by the child
can potentially affect a later OpenGL call of the type given by the
parent.  Hence, in this tree, almost every OpenGL function~$A$
is a descendant of glDraw, since there is almost always a sequence of
calls beginning with~$A$ and ending with glDraw, such that changing the
parameters of a call to~$A$, or removing call~$A$ entirely, will affect
what is displayed through glDraw.

\begin{figure}[t]
\begin{center}
\includegraphics[width=0.5\columnwidth]{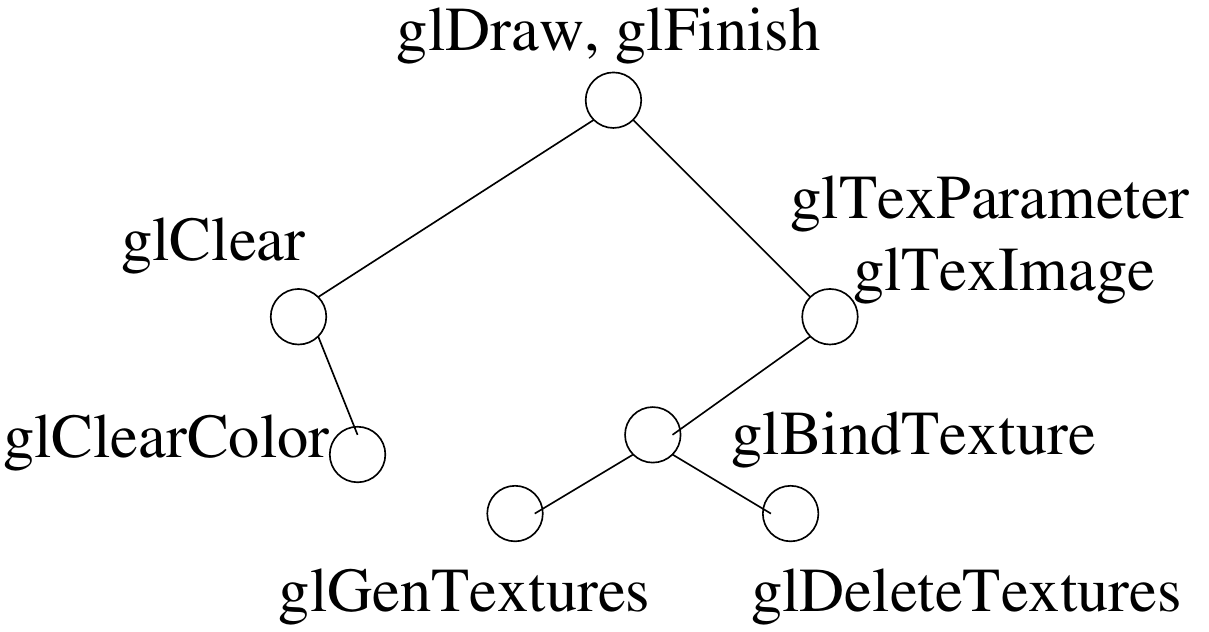}
\end{center}
\caption{\label{fig:textures} Dependency tree for some OpenGL functions.}
\end{figure}

An example of such a dependency tree is given in
Figure~\ref{fig:textures}.  The OpenGL functions that draw the next
graphics frame (glDraw), or flush the current graphics to the framebuffer
(glFinish), form the root of the dependency tree.  Note also that a call
to glGenTextures cannot affect how a call to glClearColor (setting the
``background'' color when clearing the buffer), and similarly, vice versa.

Hence, if there are two calls to glClearColor, with an intervening
call to glTexImage and no intervening calls to glDraw or glFinish,
then it is safe to prune the first call to glClearColor.
Only the second call to glColor can affect a later call to glDraw/glFinish,
and the first call to glColor cannot affect the call to glTexImage.

In general, the algorithm begins by finding the last call to
glDraw or glFinish.  No earlier call to glDraw or glFinish can
have an affect on a final call to glDraw.
Hence, we search within the log for the last OpenGL call that drew
or flushed a graphics frame.  We only prune calls until this last
drawing call.  This has the benefit of re-drawing the last graphics frame
properly upon replay, and additionally avoiding any bad corner cases
that might occur by replaying only until the middle of the commands for
a graphics frame.

After this, branches of this dependency tree are identified, such that
calls to a function within a branch may affect the calls for drawing
a graphics frame, but do not affect the behavior of any other branch.
This is illustrated in Figure~\ref{fig:textures}, where one branch
is concerned with clearing a buffer, and the other is concerned with
creating, deleting or modifying the effect of textures.

\paragraph{Textures:  a running example.}
Next, we discuss a running example, concerning textures in OpenGL~1.5.
This illustrates the general approach taken toward pruning the log
by maintaining a tree of dependencies among OpenGL calls.

A texture can be assigned to a target, such as the targets {\tt
GL\_TEXTURE\_1D}, {\tt GL\_TEXTURE\_2D}, {\tt GL\_TEXTURE\_3D}.
The {\tt glBindTexture} binds a texture~id to a target.
Other OpenGL functions operate solely on a specific target, where the target
acts as an alias for the texture~id that was most recently bound to
that target.  Thus, a call to draw a graphics frame depends only on the
texture~id that was most recently bound to a target, for each of
the targets.  Thus, for a given target, one need only retain the most
recent call to {\tt glBindTexture} for that same target.

Hence, it may {\em seem} that an older call to {\tt glBindTexture} may be
dropped from this dependency tree, if a more recent call to the same
target exists.  Unfortunately, this is not the case, due to the issue of
texture parameters.  We continue the example of textures in the following.

Each texture denoted by a texture~id has many associated parameters.
These parameters are affected both by the {\tt glTexParameter} and
the {\tt glTexImage} family of OpenGL functions.  (One can think of {\tt
glTexImage} as specifying a single texture parameter, the image associated
with a texture.  Note that these functions specify only a target, and not
a texture~id.)

In order to incorporate the dependencies on texture parameters,
we conceptually view the ``live objects'' prior to drawing a graphics
frame as consisting of {\em texture triples}\/:\hfill\newline
\hfill (texture~id, texture target, texture parameter)\hfill\newline
A function within the {\tt glTexParameter} or {\tt glTexImage} family
directly specifies a texture target and a texture parameter.
It also implicitly specifies a texture~id.  More precisely, it specifies
the texture~id corresponding to the most recent of the earlier calls
to {\tt glBindTexture} that was applied to the same
texture target.

Hence, a call within the {\tt glTexParameter} family may be dropped
if it refers to the same triple as a later call within the log
of OpenGL calls.  If a call within the {\tt glTexImage} family is
viewed as affecting an extra ``image'' parameter, then the same
rules apply.

Finally, one can now state the rule for when a call to {\tt glBindTexture}
may be dropped.  A call within the {\tt glTexParameter} or {\tt
glTexImage} family refers to a given texture triple.  This call to {\tt
glTexParameter} or {\tt glTexImage} refers implicitly to a texture~id
through an earlier call to {\tt glBindTexture}, as described earlier.
All such earlier calls to {\tt glBindTexture} must be retained,
while all other such calls may be dropped.

In the running example of textures, one can see that the size of the
log is limited to those calls relevant to a given texture triple.
Since the number of texture~ids of a graphics program will be limited,
this will determine the size of the log, in practice.

\section{Comparison: Shadow Driver versus Record-Prune-Replay}
\label{sec:comparison}

In this section, following a review of VMGL,
Section~\ref{sec:designChoices} compares the different design choices
of VMGL and the current approach.  Then Section~\ref{sec:features}
compares the features and expected performance that result from these
design choices.

\paragraph{Review of the design of VMGL.}
As mentioned earlier, VMGL~\cite{VMGL07} uses a shadow driver strategy,
similar in spirit to the earlier shadow device driver work of Swift
\hbox{et~al.}~\cite{SwiftEtAl04,SwiftEtAl06}.  In contrast, the
approach described here uses a record-prune-replay strategy.
In this section, we will always refer to the three phases
of checkpoint-restart as:  checkpoint (writing to disk a checkpoint image);
resume (resuming the original process after writing the checkpoint
image); and restart (restarting a new process, based on a
checkpoint image on disk).  In the VMGL work, they typically
refer to checkpoint as ``suspend'' and restart as ``resume''.  They
do not use the term ``restart'' at all.

Further, the VMGL work uses WireGL in its implementation.  WireGL has now
been absorbed into the current Chromium project~\cite{HumphreysEtAl02}).
We continue to refer to WireGL here, for clarity of exposition, although
any new implementation would use Chromium rather than WireGL.

Finally, in the VMGL approach, the saved graphics state for OpenGL contexts
falls into three categories.
\begin{itemize}
  \item Global context state: Including the current matrix stack, clip planes, light sources, fog setting, visual properties, etc.  
  \item Texture state
  \item  Display lists
\end{itemize}

This enabled VMGL to support most of OpenGL~1.5.

\subsection{Comparison of Design Choices}
\label{sec:designChoices}

The VMGL work is based on a shadow driver design employing
a snapshot of a virtual machine, while the the current work is based on
the use of record-prune-replay operating directly within the end user's
current desktop.  These two design choices have a variety of consequences.
Here, we attempt to codify some of the larger consequences of those
design choices.

\begin{enumerate}
  \item Virtual machine versus direct execution:
  \begin{enumerate}
    \item VMGL is designed to operate with a virtual machine such as Xen.
	At checkpoint time, VMGL uses VNC to detach a graphics viewer
	at checkpoint time, and to re-attach at restart time.  It uses
	WireGL~\cite{BuckEtAl00} to pass OpenGL commands between the
	guest operating system (containing the application) and the host
	(where the graphics are displayed).
    \item The current approach is based on DMTCP, a package for transparent
	checkpoint-restart.  A DMTCP plugin records the OpenGL calls,
	and periodically prunes the log.  The graphics application executes
	within the user's standard desktop, with no VNC.
  \end{enumerate}
  \item Checkpoint: (actions at checkpoint time)
  \begin{enumerate}
    \item VMGL calls
	on the virtual machine snapshot facility to create a checkpoint
	image (a virtual machine snapshot).
    \item The DMTCP plugin deletes the graphics window (along with
	the application's OpenGL contexts), and then calls on DMTCP to write
	a checkpoint image.
  \end{enumerate}
  \item Resume:  (resuming the original process after writing
	 a checkpoint image)
  \begin{enumerate}
    \item VMGL has very little to do on resume, and is fully efficient.
    \item The DMTCP plugin must recreate the graphics window with
	the application's OpenGL context, and then replay the log.
  \end{enumerate}
  \item Restart:  (restarting from a checkpoint image on disk)
  \begin{enumerate}
    \item VMGL restores the virtual machine snapshot along
	with its VMGL library and X~server.  WireGL connection allows
	it to re-synchronize with a stub function within a VNC viewer.
    \item The DMTCP plugin restores the
	the graphics application memory from the DMTCP checkpoint image.
	The plugin then connects to the X~server and creates a new graphics
	window and replays the pruned log, which creates a new OpenGL context,
	and restores the OpenGL state.
  \end{enumerate}
  \item {\em Restoring OpenGL ids:\/}  (In the OpenGL standard,
	many OpenGL functions return ids for numerous graphics objects,
	including textures, buffers, OpenGL programs,
	queries, shaders, etc.  Since the graphics application will
	usually cache the ids, it is important for a checkpointing package
	to use the same graphics ids on restart that were
	passed to the application prior to checkpoint.)
  \begin{enumerate}
    \item VMGL takes advantage of the fact that a new OpenGL context
	will provide ids for graphics object in a deterministic manner.
	So at restart time, VMGL makes OpenGL calls to recreate the
	graphics objects in the same order in which they were created.
    \item Unlike VMGL, The DMTCP plugin guarantees that the application
	always sees the same graphics id because the plugin only passes
	virtual ids to the application.
	The plugin provides a wrapper function around any OpenGL function
	that refers to a graphics id.  The wrapper function translates
	between a virtual id (known to the application) and a real id
	(known to the current X~server).  At restart time, the translation
	table between virtual and real ids is updated to reflect the
	real ids of the new X~server.
  \end{enumerate}
\end{enumerate}

\subsection{Comparison of Features and Performance}
\label{sec:features}

Next, we discuss some of the differences in features and performance
for VMGL and DMTCP plugin.  We {\em do not} argue that one approach is
always better than the other.  Our goal is simply to highlight what
are the different features and performance to be expected, so that
an end user may make an informed choice of which approach is best
for a particular application.

\begin{enumerate}
  \item {\em Time for checkpoint-resume:\/}  The VMGL time for
	checkpointing and resuming the original process is dominated by
	the time to write out a snapshot of the entire operating system.
	The DMTCP time for checkpointing and resuming the original process
	is dominated by two times:  the time to write out the graphics
	application memory, and the time to prune the log before checkpoint,
	and then replay the pruned log.  Note that the plugin periodically
	prunes the log, so that pruning at checkpoint time does not
	require excessive time.  (Currently, the time to replay the
	pruned log is a little larger.	This is expected to change in
	future work, when the the pruning code will be switched from
	Python to~C.)
  \item {\em Size of checkpointed data:\/}  While both approaches must
	 save the server-side data, VMGL also writes out
	 a snapshot of both the graphics application memory and its
	 entire operating system at checkpoint time.  In contrast,
	 the DMTCP plugin must write out just the graphics application.
	 However, the plugin has an extra burden in also needing to write
	 out the pruned log.
  \item {\em Rendering over a network:\/}  Since VMGL employs VNC and WireGL,
	it is trivial for VMGL to operate over a network in which the
	VNC server and the VNC viewer execute on different machines.
	(The VMGL project also makes a minor modification to VNC and
	places a VMGL-specific X~extension inside the guest operating
	system.)
	This mode is also possible for the DMTCP plugin, since DMTCP
	can checkpoint and restart a VNC server after the VNC viewer
	has disconnected.  However, most VNC servers do not support
	GPU-accelerated OpenGL implementations.  Thus, the DMTCP plugin
	approach would have to be extended to include a compatible VNC
	that uses an analog of WireGL (Chromium).  This has {\em not}
	been implemented in the current work.
  \item {\em Support for heterogeneous GPU architectures:\/}
	While both VMGL and the DMTCP plugin are vendor-independent,
	VMGL has the further advantage of being able to checkpoint
	on a computer with one type of GPU and to restart on another
	computer with a different type of GPU.  This is because the
	application always ``talks'' to the same graphics driver
	(the one saved as part of the guest virtual machine snapshot),
	and WireGL insulates this graphics driver within the guest
	from the graphics driver on the host.  One can restart on
	a different host with a different graphics driver.  In the
	case of the DMTCP plugin, see Section~\ref{sec:limitations}
	for a discussion of future work in which the plugin could
	also capture heterogeneous operation.
  \item {\em Support for distinct operating systems:\/}
	The plugin approach operates within X~Windows, and requires
	DMTCP, which has been implemented only for Linux.  The VMGL
	approach runs well over multiple operating systems, including
	Linux.  VMGL can also make use of WineD3D to translate from
	the Windows-oriented Direct3D graphics to OpenGL.
  \item {\em GPU-accelerated OpenGL graphics within a virtual machine:\/}
	The VMGL project was created specifically to provide GPU-accelerated
	graphics for a virtual machine.  Although DMTCP has been used
	to checkpoint a KVM machine~\cite{GargEtAl13}, it is not clear
	how to extend that work to support GPU-accelerated OpenGL
	for virtual machines.
  \item {\em Window manager:\/}
	Since the DMTCP plugin operates within an existing desktop,
	it is possible to restart a graphics application on a different
	desktop {\em with a different window manager}.  This is not
	possible in VMGL, since the window manager is part of the
	virtual machine snapshot, and cannot be changed on restart.
  \item {\em Virtualization of OpenGL ids:\/}
	VMGL must maintain every id ever created since the startup of
	the graphics application.  It then recreates them in the
	original order at the time of restart.	It is usually not a
	burden to maintain all ids, since typical applications create
	their graphics objects only near the beginning of the program.
	But an atypical application might choose to frequently create
	and destroy graphics ids, and this would force VMGL to carefully
	model the choice of new ids made by an OpenGL implementation
	during creation and deletion.  This model might be vendor-specific,
	thereby losing its desirable vendor-independent quality.
	In contrast, the DMTCP plugin virtualizes the OpenGL ids (see
	Section~\ref{sec:virtualization}).  This allows the plugin to
	insulate itself from the particular pattern of OpenGL ids passed
	back by the X~server.
\end{enumerate}

\section{Limitations}
\label{sec:limitations}

Several limitations are discussed.
Certain graphics operations take pointers to large data objects.
A user program may modify these data objects, or even delete
them from memory entirely.  If the data object is not available
in memory at the time of restart, then replaying the graphics
operation will not faithfully reproduce the pre-checkpoint operation.
This can be compensated for by saving a copy of the data object
at the time that the operation is executed.  But this is problematic
if the graphics operation is called repeatedly on the same data object.
A hash on the memory of the data object can be used to determine if
the data object has been modified since the last call, but this
creates significant overhead.
A good example of this occurs for the OpenGL
function  {\tt glBufferDataARB}, which enables a graphics client to
store the data associated with a server-side buffer object.

The approach presented here works well in homogeneous architectures
(same GPU and drivers on the pre-checkpoint machine and the
restart machine).  It does not support heterogeneous architectures.
This is because the graphics driver library in system memory (in RAM)
is saved with the rest of the checkpoint image.  If the checkpoint image
is transferred to another computer with a different GPU architecture,
then the original graphics driver library will fail to operate correctly.
The current approach could be generalized to heterogeneous architectures
in the future by loading and unloading appropriate graphics driver
libraries, but only at the expense of additional complication.

OpenGL~4.0 adds support for {\tt ARB\_get\_program\_binary}, which
retrieves the contents of vendor-specific program objects for later reuse.
This illustrates a situation in which any checkpointing package
for GPUs is likely to lose the ability to do heterogeneous checkpointing.

\section{Related Work}
\label{sec:relatedWork}

The current work stresses checkpointing of hardware-accelerated
OpenGL state for a standalone application.

VMGL~\cite{VMGL07} presented a pioneering approach that demonstrated
checkpointing of hardware-accelerated OpenGL state for the
first time.  That work was motivated by an approach of Swift et
al.~\cite{SwiftEtAl04,SwiftEtAl06} within the Linux kernel.
In that work, a separate shadow device driver within the kernel records
inputs to the real driver, and models the state of the real driver.
If the device driver temporarily fails due to unexpected input, the shadow
driver intercepts calls between the real device driver and the rest of
the kernel, and passes its own calls to the real device driver, in order
to place the driver back into a ``sane'' state.  This is comparable
to the approach of VMGL, except that Swift \hbox{et~al.} discuss
a shadow device driver inside the kernel, and VMGL maintains
a library within the graphics application itself to model the entire
OpenGL state (drivers and GPU hardware), and to restore that state during
restart.

The current work is distinguished from the previous two
examples~\cite{VMGL07,SwiftEtAl06} in that we do not maintain shadow
driver state, but instead we log the calls to the OpenGL API, along with
any parameters.  Both VMGL and the current work potentially suffer from
having to save the state of large data buffer objects.  Both approaches
are potentially limited when a user-space vendor-dependent library (such
as DRI) is saved as part of the application, and the application is then
restarted under another vendor's graphics hardware.  Both approaches could
unload the old library and load a new library at the time of restart.
This issue does not arise in virtual machines for which the user-space
vendor-dependent library may be eliminated through paravirtualization.

There is also a rich literature concerned with extending virtual
machine support to include run-time access from within the virtual
machine to hardware-accelerated graphics.  VMGL~\cite{VMGL07}
also appears to have been the first example in this area.
To date, aside from VMGL, none of these
approaches provide the ability to checkpoint and restore the state of
the hardware-accelerated graphics.

Several open source or free packages, including WireGL~\cite{BuckEtAl00}
(since incorporated into Chromium~\cite{HumphreysEtAl02}),
 VirtualGL~\cite{VirtualGL},
apitrace~\cite{apitrace}
provide support for extending OpenGL and other 3D
graphics APIs, so that calls to the graphics API for a
client within the guest O/S of the virtual machine are passed on to a
graphics server outside of the virtual machine.  This external graphics
server is often part of a host virtual machine, but other schemes
are available, including passing graphics commands across a network.
Schmitt \hbox{et~al.} present VirtGL~\cite{SchmittEtAl11} for virtualizing
OpenGL while preserving hardware acceleration.

Among the solutions for providing virtual access to  hardware-accelerated
GPUs are those of DowtyEtAl~\cite{DowtyEtAl09} 
(VMware Workstation and VMware Fusion), with a virtualized GPU for access
to the GPU of a host virtual machine.
Similarly, Duato et al.~\cite{DuatoEtAl10} have presented a GPU virtualization
middleware to make remote GPUs available to all cluster nodes.
Gupta \hbox{et~al.}~\cite{GuptaEtAlGVIM09} present GViM, which demonstrates
access to NVIDIA-based GPU hardware acceleration from within
the Xen virtual machine.
Along similar lines, Wegger \hbox{et~al.}~\cite{WeggerleEtAl10} present VirtGL.
Lin \hbox{et~al.}~\cite{LinEtAl10} present live migration of virtual
machines across physical machines, while supporting OpenGL.

Finally, the approach of record-prune-replay has also been applied
in the context of deterministic record-replay for a virtual machine.
In that context, one applies time slices and abstraction slices in order
to limit the record log to those calls that do not reveal sensitive
information to an unprivileged user.

\section{Experimental Evaluation}
\label{sec:experiments}

There are two larger questions that we wish to answer in this experimental
evaluation.  First, we measure the overhead of logging the OpenGL calls
(Section~\ref{sec:logging}), and also of periodically pruning the log
(Section~\ref{sec:pruning}) to prevent the log from growing too large.
These two forms of overhead are measured.

Second, in Section~\ref{sec:pruning} we measure the growth of the size
of the log over a long-running graphics program, in order to verify that
the size of the log approaches a plateau and does not grow excessively.

Finally, Section~\ref{sec:ckptRestartTimes} shows that checkpointing
requires up to two seconds, even at the larger resolutions, while
restarting requires up to 17~seconds.

While the above results are concerned with a detailed analysis of
OpenGL~1.5, Section~\ref{sec:openGL3Experiments} demonstrates that
the same approach extends to OpenGL~3.0.  That section relies
on PyMol (molecular visualization) for real-world testing.

\begin{figure}[t]
  \centering
  \hskip-10pt\includegraphics[width=0.4\columnwidth,bb=0 0 642 509]{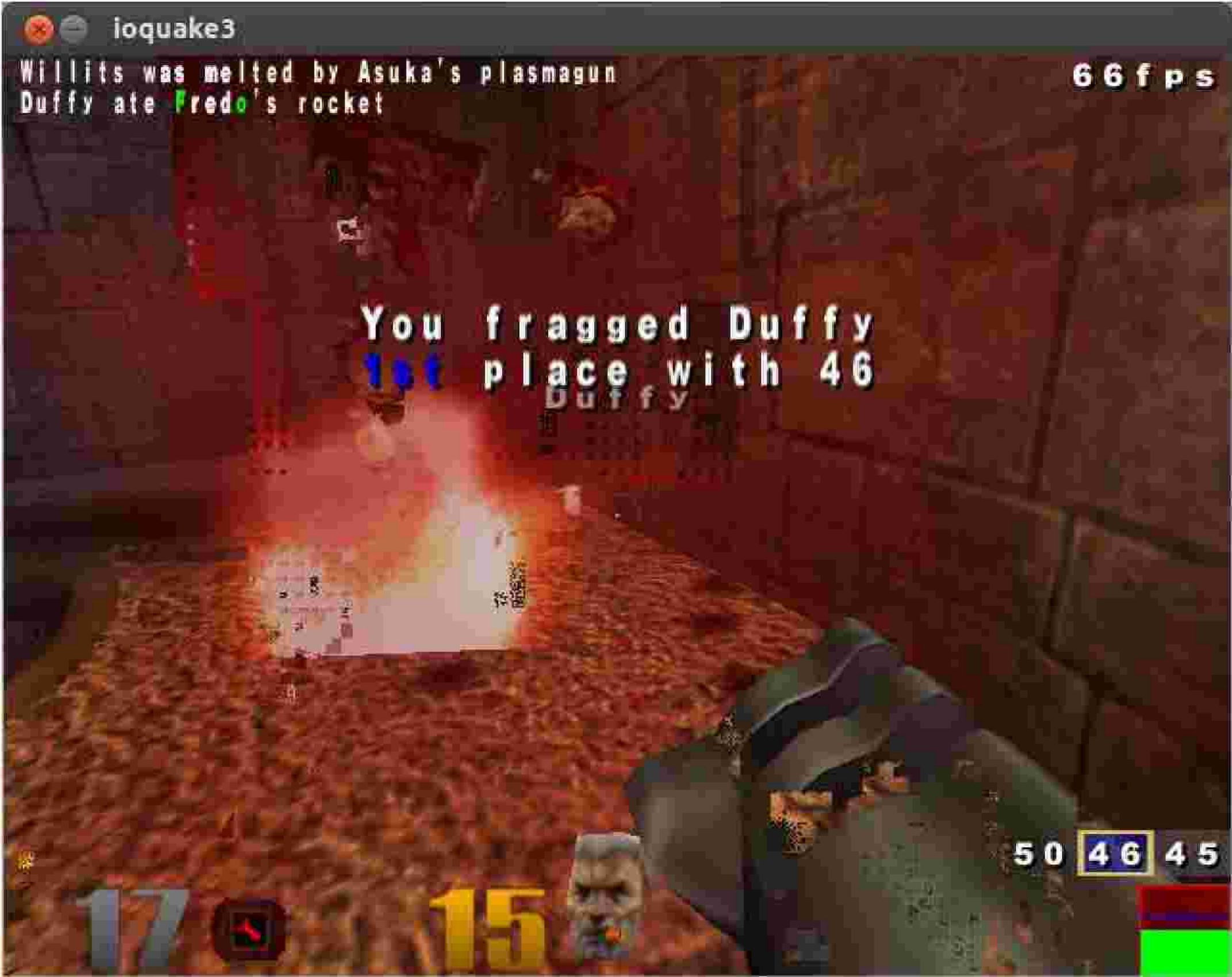}
  \caption{Ioquake running under DMTCP
  } \label{fig:ioquake-screen}
\end{figure}

The detailed testing for OpenGL~1.5 relies on ioquake3~\cite{ioquake3},
the well-known open source
game Quake3 (see Figure~\ref{fig:ioquake-screen}).
Ioquake3 is chosen due to
its existing use as an open source benchmark~\cite{videocard},
and particularly for comparability with the use of VMGL~\cite{VMGL07}.
The implementation of ioquake3 is described by Stefyn
\hbox{et~al.}~\cite{quake3-tech-report}.  For each of the experiments,
we ran the freely available 60~second Quake3 demo.

\paragraph{Configuration.}

The experiments were run on an 8-core Intel Core-i7 laptop computer with
8~GB of RAM. The host operating system was a 64-bit version of Ubuntu-13.10
with Linux kernel~3.11. For these experiments, we used an Nvidia
GT~650M graphics card and the open-source nouveau driver based on Mesa
version~9.2.1.  DMTCP-2.0 was used for the plugin implementation.

\subsection{Overhead of Logging}
\label{sec:logging}

Table~\ref{tab:performance} shows the impact on the performance of ioquake3
at different resolutions when running under DMTCP. At higher resolutions, the
time spent is dominated by the time in the GPU. In this case the overhead of
recording calls to the log is negligible.

\begin{table}[htp]
\centering
  \begin{tabular}{|c|c|c|}
   \hline
   \multicolumn{1}{|c|}{Screen} &
                         \multicolumn{1}{|c|}{FPS without} &
                         \multicolumn{1}{|c|}{FPS with}\\
   \multicolumn{1}{|c|}{Resolution} &
                         \multicolumn{1}{|c|}{DMTCP (s)} &
                         \multicolumn{1}{|c|}{DMTCP (s)}\\
   \hline
    640x480 & 70 & 90\\
    \hline
    800x600 & 70 & 90\\
    \hline
    1024x768 & 65 & 80\\
    \hline
    1600x1200 & 60 & 60\\
    \hline
    2048x1536 & 55 & 55\\
    \hline
  \end{tabular}
\caption{ \label{tab:performance}
Frames per second when running natively and under DMTCP for ioquake3}
\end{table}

The numbers in Table~\ref{tab:overhead} show the almost negligible
overhead due to the logging done by the OpenGL plugin. Three different
repeatable ``landmarks'' in the game demo were used to determine
when to checkpoint. The table shows the time to reach each of the three
checkpoints when running without DMTCP (running natively), and also the time
to reach the same checkpoint when running under DMTCP, which includes
the time to log the OpenGL calls.

\begin{table}[htp]
  \centering
  \begin{tabular}{|c|c|}
   \hline
   \multicolumn{1}{|c|}{Natively (s)} & \multicolumn{1}{|c|}{With DMTCP (s) (w/ logging)}\\
   \hline
    8 & 9\\
    \hline
    17 & 18\\
    \hline
    22 & 22\\
    \hline
  \end{tabular}
\caption{ \label{tab:overhead}                                          
Time to reach a particular point in the ioquake3 demo when running
natively and under DMTCP.
The demo was run at 640x480 resolution.}
\end{table} 

\begin{figure}[htp]
  \centering
  \includegraphics[width=0.6\columnwidth]{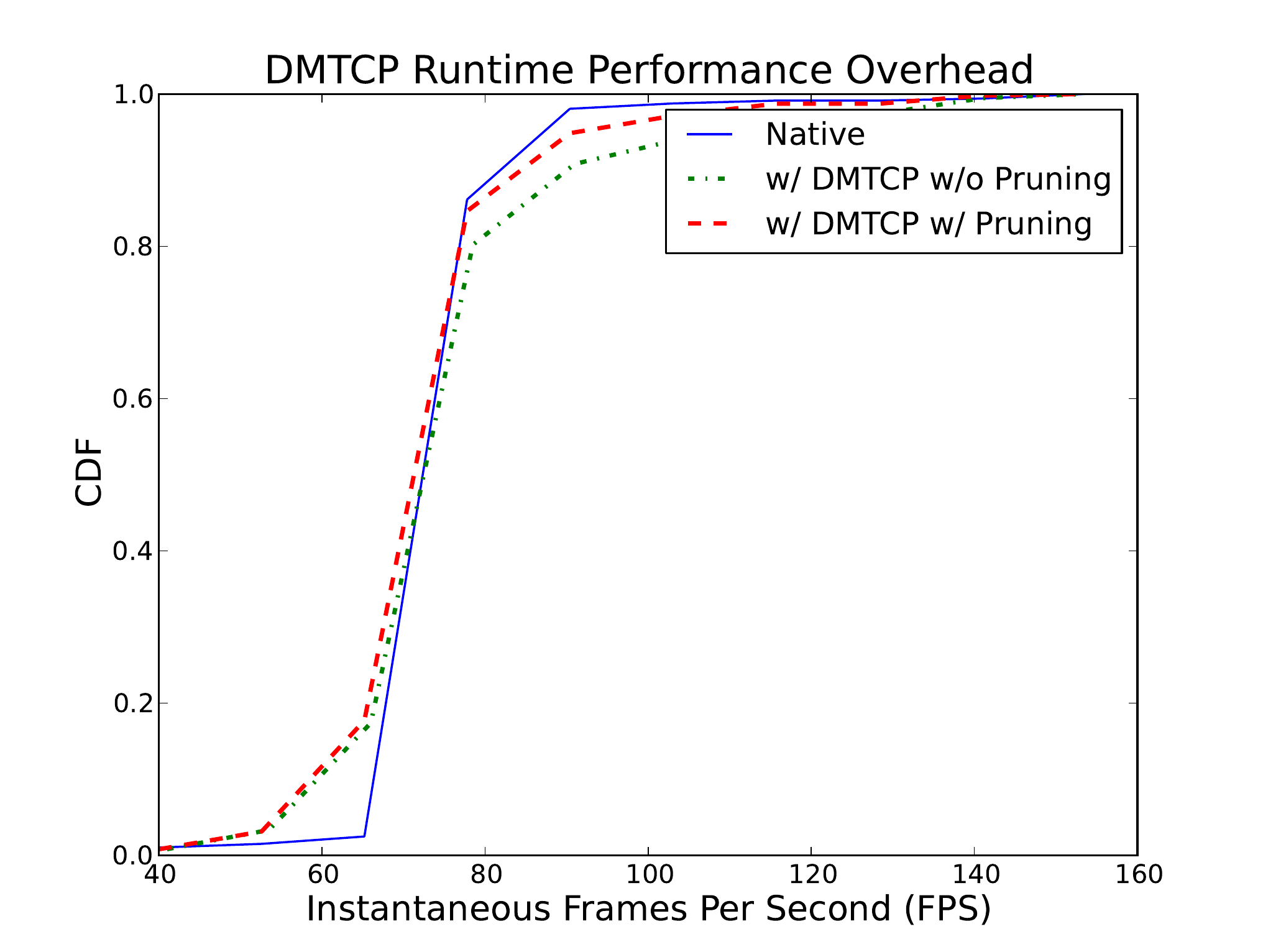}
  \caption{\label{fig:cdf-plot}
	Ioquake running under DMTCP (Instantaneous FPS \hbox{vs.} CDF).
	Jitter is small when the DMTCP curves are close to the native curve.}
\end{figure}

Figure~\ref{fig:cdf-plot} presents a plot to show the distribution
for the instantaneous number of frames per second (fps).  While
the average cumulative time with DMTCP logging is close to that of 
the native cumulative time, a cumulative distribution function can
measure the amount of jitter (delays at a particular moment)
experienced by the user.  This is important for a uniform user
experience~\cite{FunkhouserEtAl93}.
The y-axis is
the cumulative distribution function (CDF).  Ideally, the curves
when running with or without DMTCP should match closely.  The curve
shows that while the average fps for running with DMTCP is very
close to the average fps for running natively, there is a small
but significant difference for the fastest 20\% of frames and the slowest
20\% of frames.  These results are similar to those
for VMGL~\cite[Figure~4]{VMGL07}.

Figure~\ref{fig:cdf-plot} also shows that the additional CPU time
used in pruning the log does not significantly affect the overhead,
in comparison with the native running time (with no plugin).
The plugin periodically creates a child process, which prunes the log
while the parent continues to run the graphics application.  The plugin
then notifies the parent upon completion.

This experiment was run on an 8-core CPU, in a setting where
the majority of the cores were always idle.  This is expected to
be typical of future CPUs as the number of cores per CPU continues
to grow.

\subsection{Pruning: Growth of the Log}
\label{sec:pruning}

\begin{table}[htp]
\centering
\begin{tabular}{|c|c|c|c|}
   \hline
   \multicolumn{1}{|c|}{Demo} & \multicolumn{1}{|c|}{Original} & \multicolumn{1}{|c|}{Pruned} & \multicolumn{1}{|c|}{Pruning}\\
   \multicolumn{1}{|c|}{Time (s)} & \multicolumn{1}{|c|}{Log Size} & \multicolumn{1}{|c|}{Log Size} & \multicolumn{1}{|c|}{Time (s)}\\
   \hline
    5 & 18 MB & 133 KB & 1.3\\
    \hline
    10 & 28 MB & 139 KB & 2.2\\
    \hline
    20 & 64 MB & 145 KB & 4.9\\
    \hline
    30 & 100 MB & 155 KB & 7.8\\
    \hline
    45 & 158 MB & 163 KB & 12.0\\
    \hline
    60 & 190 MB & 164 KB & 15.2\\
    \hline
    75 & 210 MB & 163 KB & 17.7\\
    \hline
    90 & 287 MB & 165 KB & 22.1\\
    \hline
    105 & 352 MB & 164 KB & 28.5\\
    \hline
    120 & 433 MB & 163 KB & 34.3\\
    \hline
  \end{tabular}
\caption{ \label{tab:prune}
Size of pruned log and the time to prune for increasing demo time. The demo was
running at 640x480 resolution.}
\end{table} 

Table~\ref{tab:prune} shows the growth in the size of the pruned log and
the time to prune the log.  These statistics are taken as a function
of the time spent in the ioquake game demo.  While the
size of the original log grows 9~times, the size of the pruned
log grows just~26\%.  More important, the size of the pruned log
reaches an asymptotic plateau.

The 17.7~seconds to prune the log after 45~seconds
represents an overhead of 40\%.  This excessive overhead is due to
our current implementation in which logged calls are represented
as strings similar to the format used in source code.  Further, the
pruning is currently executed by a Python program (slightly optimized
using Cython).  A future implementation will use a binary encoding,
and pruning will be done in a compiled language for efficiency.
Section~\ref{sec:log} further describes the rationale for currently
representing the log as a human-readable log based on ASCII text in a
file on disk.  In the current implementation, the overhead of
pruning the log is reduced to almost nothing by running that code in a child
process, which typically runs on a separate CPU core.

\begin{figure}[htp]
  \centering
  \includegraphics[width=0.5\columnwidth]{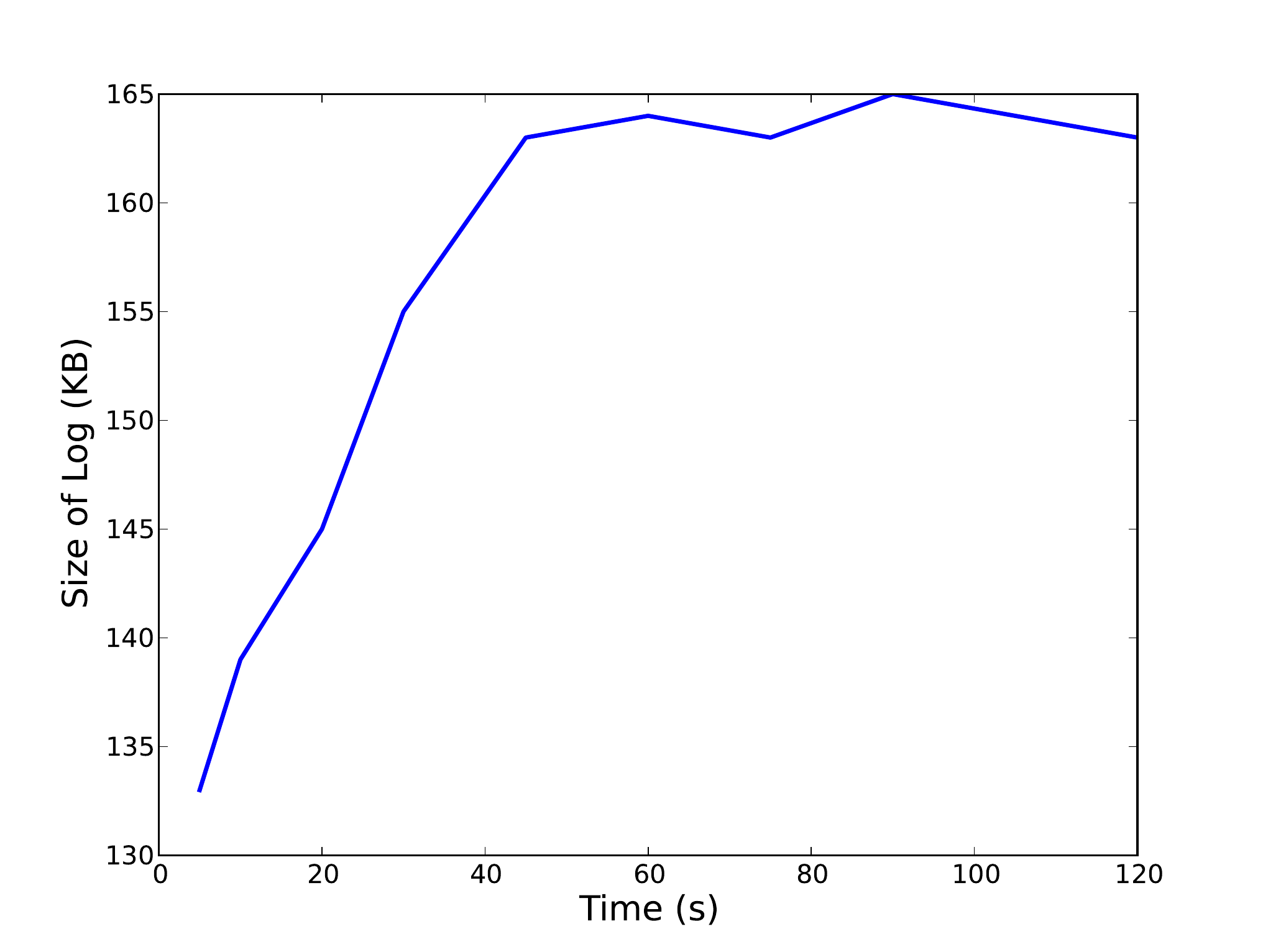}
  \caption{\label{fig:prune-plateau}
	The size of the log eventually reaches a plateau, and does
	not grow further.}
\end{figure}

\subsection{Checkpoint-Restart Times}
\label{sec:ckptRestartTimes}

Figure~\ref{fig:prune-plateau} graphically demonstrates that the
size of the log reaches a plateau (approximately 163~KB in this case)
and does not grow further as the demo continues to execute. 

Table~\ref{tab:scalability} shows the variation in image size, the time to
checkpoint, and the time to restart at different resolutions for ioquake3. The
checkpoints were taken at arbitrary points while running the Quake3 demo. The
time to restart includes the time taken to prune the log file.

\begin{table}[htp]
\centering
\begin{tabular}{|c|c|c|c|}
   \hline
   \multicolumn{1}{|c|}{Screen} & \multicolumn{1}{|c|}{Image} & \multicolumn{1}{|c|}{Ckpt} & \multicolumn{1}{|c|}{Restart}\\
   \multicolumn{1}{|c|}{Resolution} & \multicolumn{1}{|c|}{Size} & \multicolumn{1}{|c|}{Time (s)} & \multicolumn{1}{|c|}{Time (s)}\\
   \hline
   640$\times$480 & 32 MB & 1.5 & 15\\
    \hline
    800$\times$600 & 32 MB & 1.7 & 16\\
    \hline
    1024$\times$768 & 33 MB & 1.8 & 17\\
    \hline
    1600$\times$1200 & 33 MB & 2.0 & 17\\
    \hline
  \end{tabular}
\caption{ \label{tab:scalability}                                          
Scaling of the image size, checkpoint, and restart time with DMTCP for ioquake}
\end{table}

\subsection{Tests on PyMol (using OpenGL~3.0)}
\label{sec:openGL3Experiments}

\begin{figure}[htp]
  \hskip120pt\includegraphics[width=0.5\columnwidth,bb=0 0 862 527]{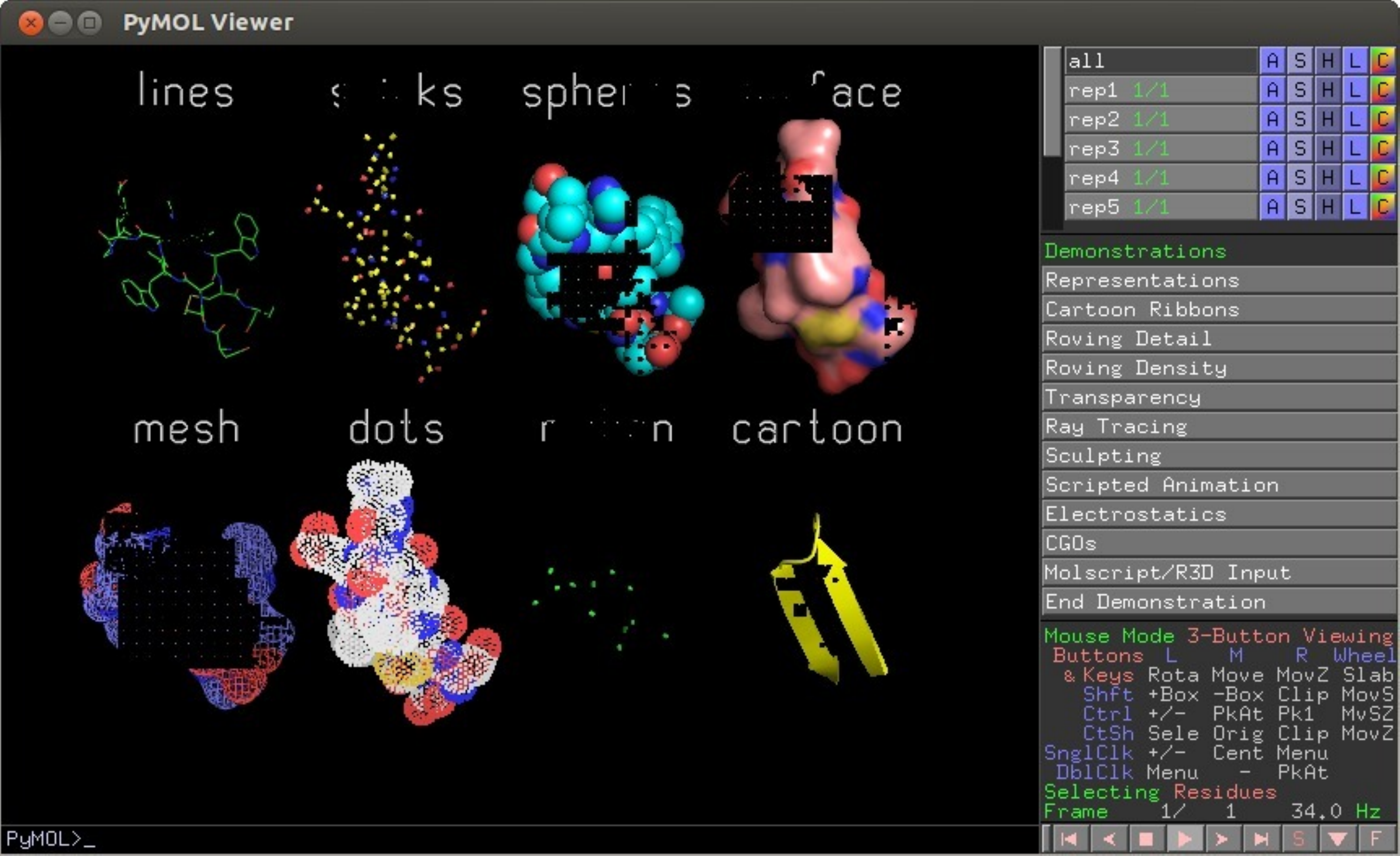}
  \caption{\label{fig:pymol} PyMol screenshot}
\end{figure}

To test the proposed approach with OpenGL~3.0 we used a widely-used
open-source molecular visualization software, PyMOL~\cite{PyMol}.
Figure~\ref{fig:pymol} shows the application running a demonstration with
different representations of a protein. We observe that the application runs
with a 9\% overhead on the frames-per-second because of the DMTCP plugin. The
time to checkpoint, and the time to restart (that includes replaying of the
logs) are 1.4~seconds, and 10~seconds, respectively. The performance overhead,
and the restart times are better because of the OpenGL~3.0 features calls
to shader processing units, which do more work for the same
log entry by the plugin.

\section{Future Work}
\label{sec:futureWork}

A production-quality version of this software will be contributed
to the DMTCP open source project for checkpoint-restart.  In that version,
the log of OpenGL calls will be stored in a binary format, instead
of using ASCII strings.  Two other enhancements will make it still
faster.  First, the program for pruning the log will be re-written
in C, and made into a software module in the main executable.
Second, the log will be maintained in RAM, instead of being stored
as a file on disk.

Next, support for heterogeneous GPU architectures will be considered,
in which the GPU on the restart machine is different from the
GPU on the pre-checkpoint machine.  See Section~\ref{sec:limitations}
for a further discussion of this issue.

Since software such as WireGL can be used as part of a larger
rendering farm, and since DMTCP directly supports checkpointing of
networks~\cite{AnselEtAl09}, this work can be extended to support
checkpointing over a cluster of GPU-enabled workstations for
visualization.

The work of VMGL~\cite{VMGL07} has already demonstrated the
 possibility of using WireGL~\cite{BuckEtAl00} and a VNC implementation
to render the 3D~graphics inside a virtual machine through transport
to a GPU-enabled host.  The same scheme could be applied in this approach.
While VNC had used WireGL and TightVNC, it is currently possible to
use either WireGL with a choice of VNC implementations, or else
to use VirtualGL~\cite{VirtualGL} and its recommended pairing with TurboVNC.

Support for Direct3D under Windows will be considered by running
Windows inside a virtual machine, and making use of the
WineD3D~\cite{WineD3D} library for translation of Direct3D to OpenGL.

The approach of this work will be extended to work with Xlib. This will
eliminate the need for supporting GLUT and SDL. Furthermore, this will
directly support all 2D~applications based on Xlib. This improves on
the current practice of using VNC to support checkpointing of 2D~graphics.

\section{Conclusion}
\label{sec:conclusion}

This work shows feasibility for a record-prune-replay approach to
checkpointing 3D~graphics for both OpenGL~1.5 and OpenGL~3.0.
Checkpointing requires up to two seconds.
While the current run-time overhead is high (for example, 40\% overhead
for periodically pruning the log), this is attributed to the current
inefficient representation of OpenGL calls in the log as ASCII strings
in a file on disk.
A binary data representation is planned for the future.
This will eliminate the need to do pruning in a second CPU core.

Previously, the only approach to transparently saving the state of
GPU-accelerated OpenGL as part of checkpoint-restart was to use VMGL's
shadow device driver along with a virtual machine snapshot.
That approach was demonstrated for OpenGL through 78,000 lines of code.
As an indication of the relative levels of effort, the current approach
required only 4,500 lines of code for OpenGL~1.5 and and 6,500 lines
in total.

The current approach is implemented as a plugin that extends the
functionality of the DMTCP checkpoint-restart package.  This allows the
graphics programs to be checkpointed directly, without the intervention
of a virtual machine and its snapshot capability.

\section*{Acknowledgments}

We would especially like to thank James Shargo for demonstrating that
apitrace supports a record-replay approach, and then suggesting
that although he did not personally have the time, we should pursue
a record-prune-replay approach for checkpoint-restart.  We would also
like to thank Andr\'es Lagar-Cavilla for his earlier discussions on the
design and implementation of VMGL.  Finally, we would also like to thank
Daniel Kunkle for his insights during an earlier investigation into the
complexities of checkpointing OpenGL in~2008, at a time when much of
the software landscape was less mature.

\bibliographystyle{acm}

\end{document}